\documentclass[english,american,aps,preprint]{revtex4}
\usepackage[T1]{fontenc}
\usepackage[latin9]{inputenc}
\usepackage{amsmath}
\usepackage{amssymb}
\usepackage{graphicx}

\makeatletter
\@ifundefined{textcolor}{}
{%
 \definecolor{BLACK}{gray}{0}
 \definecolor{WHITE}{gray}{1}
 \definecolor{RED}{rgb}{1,0,0}
 \definecolor{GREEN}{rgb}{0,1,0}
 \definecolor{BLUE}{rgb}{0,0,1}
 \definecolor{CYAN}{cmyk}{1,0,0,0}
 \definecolor{MAGENTA}{cmyk}{0,1,0,0}
 \definecolor{YELLOW}{cmyk}{0,0,1,0}
}

\makeatother

\usepackage{babel}
\begin{document}

\preprint{BROWN-HET-1636, NORDITA-2012-90, RH-10-2012}

\title{Black holes without firewalls}

\author{Klaus Larjo}

\email{klaus_larjo@brown.edu}

\address{Department of Physics, Box 1843, Brown University, Providence, RI,
02912, USA}

\author{David A. Lowe}

\email{lowe@brown.edu}

\address{Department of Physics, Box 1843, Brown University, Providence, RI,
02912, USA}

\author{Larus Thorlacius}

\email{larus@nordita.org}

\address{Nordita, KTH Royal Institute of Technology and Stockholm University,
Roslagstullsbacken 23, SE-106 91 Stockholm, Sweden\\{\rm and}\\ University
of Iceland, Science Institute, Dunhaga 3, IS-107 Reykjavik, Iceland}
\begin{abstract}
The postulates of black hole complementarity do not imply a firewall
for infalling observers at a black hole horizon. The dynamics of the
stretched horizon, that scrambles and re-emits information, determines
whether infalling observers experience anything out of the ordinary
when entering a large black hole. In particular, there is no firewall
if the stretched horizon degrees of freedom retain information for
a time of order the black hole scrambling time.
\end{abstract}
\maketitle

\section{Introduction: complementarity or firewall?}

Black hole complementarity was introduced in \cite{Susskind:1993if}
in terms of three postulates for black hole evolution:
\begin{enumerate}
\item The process of formation and evaporation of a black hole, as viewed
by a distant observer, can be described entirely within the context
of standard quantum theory. In particular, there exists a unitary
S-matrix which describes the evolution from infalling matter to outgoing
Hawking-like radiation.
\item Outside the stretched horizon of a massive black hole, physics can
be described to good approximation by a set of semi-classical field
equations.
\item To a distant observer, a black hole appears to be a quantum system
with discrete energy levels. The dimension of the subspace of states
describing a black hole of mass $M$ is the exponential of the Bekenstein
entropy $S(M)$.
\end{enumerate}
These postulates refer to observations made outside the black hole
and provide a basis for a phenomenological description that is consistent
with unitarity. A further key assumption, based on the equivalence
principle, was made in \cite{Susskind:1993if} and can be expressed
as a fourth postulate that applies to observers who enter the black
hole:

\begin{enumerate}
\item[4]
An observer in free fall experiences nothing out of the ordinary upon crossing the horizon of a large black hole.
\end{enumerate}The combination of this assumption and the three original postulates
requires one to give up the notion of spacetime locality. In particular,
the fate of observers entering a large black hole is very different
depending on the frame of reference: In their own rest frame they
pass unharmed through the horizon and only come to harm as they approach
the curvature singularity, while from the viewpoint of distant observers
they never pass through the horizon at all but are instead absorbed
into the stretched horizon and thermalized before being re-emitted
along with the rest of the black hole in the form of Hawking radiation.
It was argued in \cite{Susskind:1993mu} that no low-energy observer
can detect violations of known laws of physics even if information
carried by infalling matter appears to be duplicated in the outgoing
Hawking radiation. 

The stretched horizon is a surface outside the global black hole horizon
that remains timelike. Outside observers ascribe non-trivial microphysical
dynamics to the stretched horizon that serves to absorb, thermalize,
and eventually re-emit the information contained in infalling matter.
The usual thermodynamics of black holes is assumed to arise from a
coarse graining of this (unspecified) microscopic dynamics. From the
point of view of outside observers, no information ever enters the
black hole in this description and the stretched horizon is the end
of the road for all infalling matter. In that sense it is indeed a
firewall. According to the fourth postulate the story is very different
for an infalling observer. The spacetime curvature is weak at the
horizon of a large black hole and an infalling observer should not
notice anything out of the ordinary upon crossing the horizon. In
a recent paper Almheiri \emph{et al.} \cite{Almheiri:2012rt} claim,
however, that the first two postulates imply that an infalling observer
must also see a firewall %
\footnote{Closely related earlier work appeared in \cite{braunstein}.%
}. In other words, that the fourth postulate is inconsistent with the
others.
\begin{figure}
\includegraphics[scale=0.75]{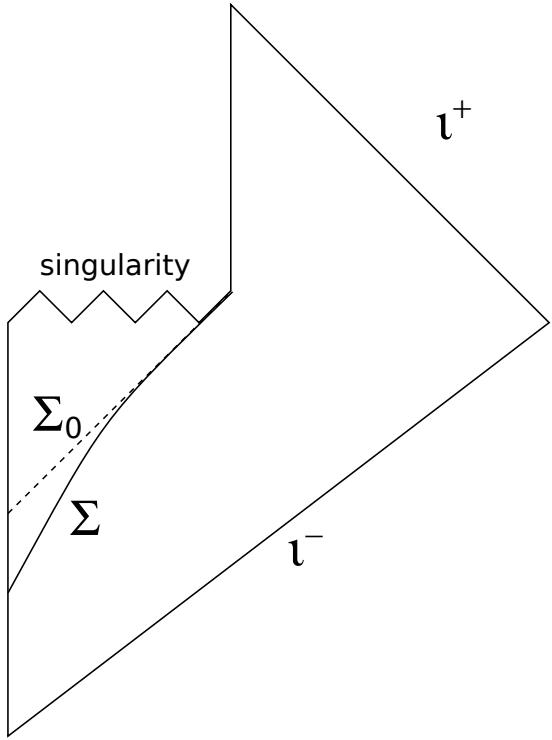}

\caption{\label{fig:Penrose-diagram-for}Penrose diagram for black hole evaporation.
$\Sigma_{0}$ is the global horizon and $\Sigma$ is a stretched horizon.}

\end{figure}

The microscopic stretched horizon in \cite{Susskind:1993if} was placed
at a proper distance of order the Planck length away from the global
horizon. More generally in the present work, we require the stretched
horizon be placed at some large fixed redshift from asymptotic infinity
\cite{Thorne:1986iy}. According to the second postulate, physics
outside the stretched horizon is described by semi-classical field
equations of some low-energy local effective field theory. The definition
of a low-energy theory includes specifying a cut-off and in the black
hole context this means that the spatial slices, on which the effective
theory is defined, terminate at an \emph{effective }stretched horizon
located outside the microscopic stretched horizon. For concreteness,
let us consider a quasi-static spherically symmetric black hole as
shown in figure \ref{fig:Penrose-diagram-for}. The effective stretched
horizon can then be taken as the surface $\Sigma$, where fiducial
observers (who remain at rest with respect to the black hole) would
measure a local temperature equal to a cut-off scale. Equivalently,
$\Sigma$ is the surface such that a radially outgoing massless particle
is red-shifted from the cut-off energy at $\Sigma$ to the characteristic
energy of Hawking radiation at infinity. In this approach, anything
that is inside the effective stretched horizon is represented by degrees
of freedom living on the effective stretched horizon. This includes
the entire black hole region and the region between $\Sigma$ and
the global horizon, as indicated in figure \ref{fig:Penrose-diagram-for}.
If the cut-off energy is taken very high, close to the Planck energy,
then $\Sigma$ approaches the microscopic stretched horizon of \cite{Susskind:1993if}.
For lower values of the cut-off, the dynamics on the effective stretched
horizon is in principle obtained from the dynamics on the underlying
microscopic horizon by renormalization. 

The argument of \cite{Almheiri:2012rt} proceeds as follows: At very
late times we have by supposition a pure state consisting only of
outgoing Hawking radiation. Since we know the laws of physics up to
the stretched horizon we can evolve this state back mode by mode from
late times to the stretched horizon. The authors of \cite{Almheiri:2012rt}
then seem to introduce the hidden assumption that even beyond the
surface $\Sigma$ we can still evolve the mode back all the way to
the global horizon, ignoring the stretched horizon degrees of freedom.
At that point the argument can be reduced to that of a single mode,
since by locality such a mode very close to the global horizon does
not have time to entangle with the stretched horizon, and thus cannot
entangle with other outgoing Hawking modes emitted at later times.
Thus, once one has evolved this single late-time mode back to a point
very close to the global horizon, unitarity and local quantum field
theory prevent any entanglement between the outgoing mode and the
black hole state. This is sufficient to guarantee that an infalling
observer will see high energy modes. In fact, the states obtained
in this manner are finite excitations of the so-called Boulware vacuum,
which is well-known to have a divergent stress energy tensor on the
horizon \cite{Candelas:1980zt}. Vacuum states regular on the horizon
include the Hartle-Hawking vacuum or Unruh vacuum \cite{Candelas:1980zt},
which requires entanglement between outgoing Hawking modes and negative
energy modes inside the black hole.

The firewall argument of \cite{Almheiri:2012rt} is flawed because
the stretched horizon has been dispensed with. In the effective field
theory description of Postulate 2, Hawking radiation is emitted from
the stretched horizon, which is a boundary of the spacetime. The fact
that at late times the state of the stretched horizon is maximally
entangled with the early Hawking radiation is no more of a problem
than the corresponding statement about the remaining embers of a burning
lump of coal that started out in a pure state. The firewall problem
only arises if one attempts to extend the semiclassical description
to the region inside the stretched horizon. If there is no entanglement
between the outgoing modes and the state of the black hole, as is
argued in \cite{Almheiri:2012rt}, then the state of the field is
given by an excitation of the Boulware vacuum, which has a stress
energy tensor that is power-law divergent as a function of proper
distance as the global horizon is approached \cite{Candelas:1980zt}.
In that case, infalling observers encounter drama already outside
the stretched horizon, in violation of black hole complementarity.
We will give a counter-example below, where the semiclassical description
outside the stretched horizon is compatible with unitarity and locality
and the expectation value of the stress energy tensor remains finite
in that region. This is achieved by making a different assumption
about the semiclassical state of the system. Our semiclassical construction
involves a firewall but only inside the stretched horizon, where the
effective field theory of Postulate 2 no longer applies. The presence
of a firewall, both in our example and in {[}3{]}, is at odds with
Postulate 4 but this is hardly surprising. Black hole complementarity
was after all put forward to address problems arising from applying
semiclassical theory in a region extending inside the stretched horizon.

Observations made by infalling observers are only well described in
the local effective theory of Postulate 2 as long as they remain outside
the surface $\Sigma$ in figure \ref{fig:Penrose-diagram-for} and
not after they pass through it. An alternative description should
be possible, involving an effective quantum field theory on time slices
where an infalling observer has low energy in the local frame of the
slice \cite{Lowe:1995ac,Lowe:1995pu}. However, to describe an infalling
observer crossing the global horizon of the black hole requires time
slices that extend past the location of the stretched horizon in the
original effective field theory, making it difficult to map states
and observables from one low-energy theory to the other. Moreover,
it has been argued that, due to the large relative boosts involved,
the effective low-energy description on time slices, such that both
an infalling observer who has entered the black hole and the outgoing
Hawking radiation are at low energy, cannot be a local field theory
\cite{Lowe:1995ac,Lowe:1995pu,Lowe:1999pk,Giddings:2004ud}. 

With some new assumptions about the stretched horizon dynamics, and
taking care with the application of the semiclassical approach, we
will argue in the following section that information may be recovered
without introducing a firewall for infalling observers. Various alternatives
or modifications of the firewall scenario have appeared in \cite{Nomura:2012sw,Mathur:2012jk,Chowdhury:2012vd,Avery:2012tf,Banks:2012nn,Giddings:2012dh,Hwang:2012nn,Hossenfelder:2012mr}
but for the most part these are also alternatives to black hole complementarity.
The argument in the present paper, on the other hand, is consistent
with black hole complementarity, as formulated in \cite{Susskind:1993if},
with additional assumptions about the dynamics of the stretched horizon.

\section{Emergence of information}

Let us begin by revisiting the setup of Hayden and Preskill \cite{Hayden:2007cs},
making explicit some of the relevant timescales. Alice throws her
diary into an old black hole, where more than half the entropy has
been emitted in Hawking radiation, and Bob measures the outgoing Hawking
quanta, having first faithfully recorded all the quanta previously
emitted by the black hole. That Bob can manipulate the state of the
Hawking radiation in a relatively short time can be argued as follows.
On average, a black hole of mass $M$ emits a Hawking particle every
$M$ units of time, with an average energy of $\frac{1}{M}.$ Hence
the total number of emitted quanta during the lifetime of a black
hole will be $M^{2}$, and the total lifetime will be $M^{3}.$ This
timescale can be thought of as $M^{2}$ 'boxes' of length $M$, and
distributing the emission times of the $M^{2}$ Hawking particles
into these 'boxes' gives Bob access to roughly

\[
N\sim\left(M^{2}\right)^{M^{2}}
\]
different states; more than enough to differentiate between the $e^{M^{2}}$
microstates making up the black hole. 

As discussed in \cite{Hayden:2007cs,Sekino:2008he}, the scrambling
time of a black hole is given by
\[
t_{\mathrm{scramble}}\sim M\,\log\, M.
\]
In this time an average of $\log\, M$ Hawking particles will be emitted,
leading to a possible problem: since these Hawking particles can be
entangled with the diary before the black hole scrambles, there may
be a modification of the high frequency quanta, and hence an infalling
observer may see a firewall. On the other hand, as argued in \cite{Hayden:2007cs},
if the entanglement only appears after $t_{scramble}$ one finds compatibility
with the postulates of black hole complementarity.

One can use information theory arguments to place a lower bound on
the time scale of information retrieval. This is computed in \cite{Hayden:2007cs}
in eqn (1). They find the probability for failure to decode a $k$-bit
message in the diary satisfies
\begin{equation}
P_{fail}\leq2^{k}2^{-s}\label{eq:pfail-1}
\end{equation}
where $s$ is the number of bits that Bob reads after the diary is
thrown in. Now this seems to imply the information comes out faster
than the scrambling time if $k\approx1$, from which one might infer
a firewall. There is, however, an error in this train of logic, since
the authors of \cite{Hayden:2007cs} assume scrambling has already
happened when they make the estimate in \eqref{eq:pfail-1}. Prior
to scrambling the emission rate of quantum information is not governed
by \eqref{eq:pfail-1} but rather depends on details of the stretched
horizon dynamics.%
\footnote{If, for example, the message was unentangled with the bits received
by Bob, he would still have a probability of success of $2^{-k}$
of decoding the message by measuring the first $k$ bits. However
this does not improve as additional bits are measured, so does not
correspond to a useful measurement of information in the diary.%
}

Let us try to model these effects in more detail to determine their
implications for the stretched horizon theory. Consider an old black
hole prior to the diary being thrown in. It is fully entangled with
the train of Hawking radiation that has already been emitted, and
its state can be written as 
\[
|\Psi\rangle_{BBh}=\sum_{i}c_{i}|i\rangle_{B}\otimes|i\rangle_{Bh},
\]
with $i\in[1,N]$ indexing the basis states of the black hole, and
$|i\rangle_{B}$ being the corresponding string of Hawking radiation
recorded by Bob. Tracing over the Hawking radiation, the black hole
density matrix is diagonal, with uniform entries due to the maximal
entanglement, 
\[
\rho_{B}=\sum_{i}|c_{i}|^{2}|i\rangle_{Bh}\langle i|_{Bh}=\frac{1}{N}\mathbb{I}_{N},\qquad\mathrm{with}\quad N=\exp\left(M^{2}\right).
\]
Into this state Alice throws her diary, which we initially take to
consist of $k$ bits in a pure state. Without loss of generality,
we can take the state to be $|+_{1},\ldots,+_{k}\rangle_{A}\equiv|(+)_{k}\rangle_{A}.$
Immediately after the diary is inside the black hole, but not yet
scrambled, the state and the black hole density matrix are given by
\begin{equation}
\begin{aligned}|\Psi\rangle & =\sum_{i}c_{i}|i\rangle_{B}\otimes|i,(+)_{k}\rangle_{BhA},\\
\rho_{BhA} & =\frac{1}{N}\left(\begin{array}{cccc}
\mathbb{I}_{N} & 0 & \cdots & 0\\
0 & 0 & \cdots & 0\\
\vdots & \vdots & \ddots & 0\\
0 & 0 & 0 & 0
\end{array}\right),
\end{aligned}
\label{eq:minentang}
\end{equation}
where the density matrix $\rho$ is a $2^{k}N\times2^{k}N$ matrix,
written in term of $N\times N$ blocks above. The vanishing blocks
of $\rho_{BhA}$ correspond to states involving $|-\rangle_{A}$,
over which Alice's diary does not have support yet. This is to be
compared with the maximally entangled $2^{k}N\times2^{k}N$ matrix
\begin{equation}
\rho_{max}=\frac{1}{2^{k}N}\left(\begin{array}{cccc}
\mathbb{I}_{N} & 0 & \cdots & 0\\
0 & \mathbb{I}_{N} & \cdots & 0\\
\vdots & \vdots & \ddots & 0\\
0 & 0 & 0 & \mathbb{I}_{N}
\end{array}\right).\label{eq:maxentang}
\end{equation}

In modeling the stretched horizon dynamics, we assume that there is
one degree of freedom per unit Planck area, consistent with the third
postulate. The transverse wavelength of modes emitted from the stretched
horizon then ranges from of order $M$, for the low-angular momentum
modes that make up the bulk of the Hawking radiation that reaches
distant observers, to of order one in Planck units, for high-angular
momentum modes that never emerge far from the black hole and are re-absorbed
by the stretched horizon. 

A simple model for a long transverse wavelength, Hawking particle
emitted from the stretched horizon is an operator close to the identity
operator in the $2^{k}N\times2^{k}N$ dimensional Hilbert space, acting
on all the stretched horizon states with approximately equal weight
\[
\mathcal{O}_{long}=\mathbb{I}{}_{2^{k}N}+\epsilon
\]
where $\epsilon$ is some small perturbation with vanishing trace.
We see in each case \eqref{eq:minentang} and \eqref{eq:maxentang}
that 
\begin{equation}
\mathrm{Tr}\rho\times\mathcal{O}_{long}=1\label{eq:oplong}
\end{equation}
so these operators do a good job of making the emitted radiation look
thermal, regardless of the stretched horizon state. 

On the other hand, a model for the emission of a short transverse
wavelength mode would be to pick an operator such as
\begin{equation}
\mathcal{O}_{short}=\left(\begin{array}{cccc}
\mathbb{I}_{N} & 0 & \cdots & 0\\
0 & 0 & \cdots & 0\\
\vdots & \vdots & \ddots & 0\\
0 & 0 & 0 & 0
\end{array}\right)\label{eq:localop}
\end{equation}
In this case, immediately after the diary was thrown in, we would
find 
\begin{eqnarray*}
\mathrm{Tr}\rho_{BhA}\times\mathcal{O}_{short} & = & 1\\
\mathrm{Tr}\rho_{max}\times\mathcal{O}_{short} & = & 1/2^{k}
\end{eqnarray*}
so we see the answer is highly sensitive to the state of the stretched
horizon. The danger is that operators of the form \foreignlanguage{english}{$\mathcal{O}_{short}$}
will lead to strong modification of the high-frequency near-horizon
Hawking modes, giving rise to a firewall. For this to happen, the
short transverse wavelength modes coupling to such operators would
have to themselves scramble and become entangled with the early Hawking
radiation on a timescale that is short compared to the black hole
scrambling time $t_{scramble}$. If, however, the stretched horizon
dynamics is causal, then short transverse wavelength modes are unable
to scramble (i.e. achieve approximate global thermalization with respect
to the measure described in \cite{Hayden:2007cs}) in a time $M\log M$.
The fastest a localized signal can causally traverse the stretched
horizon is in time $M$ using time measured at the stretched horizon.
This then redshifts to $M^{2}$ when measured using Schwarzschild
time. We therefore introduce another new assumption about the dynamics
of the stretched horizon theory -- that it be local and causal. Without
this assumption the stretched horizon dynamics can in principle contaminate
the causal physics outside, violating Postulate 2.

The bounds on information retrieval time placed in \cite{Hayden:2007cs}
are lower bounds. The diary will scramble most efficiently if it is
coded into modes with transverse wavelength of order $M$, as exemplified
by the operator \eqref{eq:oplong}. Due to the long transverse wavelength,
such modes couple globally to the stretched horizon degrees of freedom,
and there is no causal bound preventing an $M\log M$ scrambling time.
If, on the other hand, the information in the diary is present in
short transverse wavelength modes, such as \eqref{eq:localop} then
it may be emitted more slowly, on a timescale of order $M^{2}$ or
longer.

It is, however, necessary to assume there is a genuine information
retention time during which no ``prompt'' information is emitted
from the stretched horizon while these long transverse wavelength
modes scramble. This distinguishes the dynamics of the stretched horizon
from, for instance, an accelerating mirror which would indeed look
like a firewall from the point of view of a freely falling observer.
The dynamics of the stretched horizon must be such that the reflection
coefficient vanishes, the information is retained for a time $t_{scramble}$,
at which point it is then primarily emitted in long transverse wavelength
modes.

It is important in the argument of \cite{Hayden:2007cs} that the
diary be much smaller than the black hole. This can also be seen from
the following estimate of the maximum number of degrees of freedom
that can scramble fast enough. We ask that a causal signal from a
cell of size $\lambda_{min}$ on the stretched horizon overlaps with
a neighboring cell after $t_{scramble}$$ $ and assume that this
is sufficient for scrambling, with respect to the measure of \cite{Hayden:2007cs},
to take place. This is rather strong assumption about the efficiency
of the scrambling dynamics so the resulting number of fast scramblers
is likely to be an overestimate. The above condition implies $\lambda_{min}\sim\log M$,
in which case the number of independent fast scrambling degrees of
freedom is of order $\left(M/\log M\right)^{2}$. Sending in a larger
diary than this will compromise the rapid rate of information retrieval,
as more generic short transverse wavelength modes scramble on a slower
timescale of order $M^{2}$.
\begin{figure}
\includegraphics[scale=0.5]{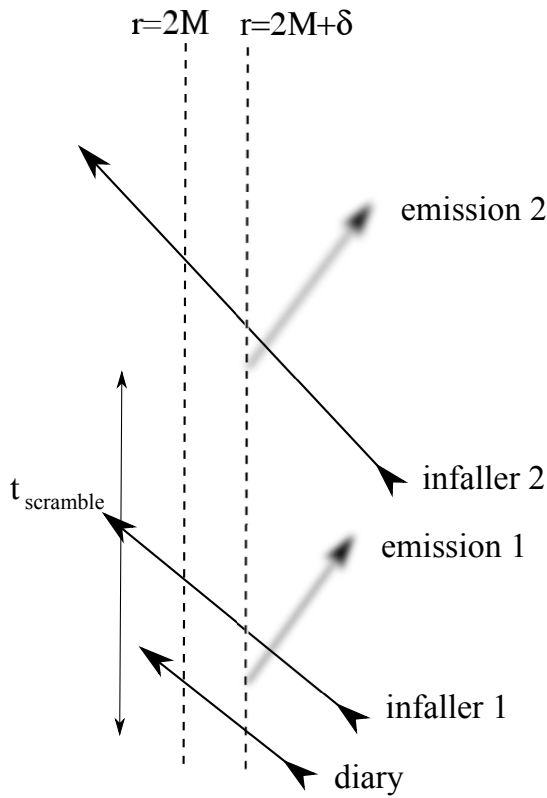}\caption{\label{fig:Different-infalling-observers}Different infalling observers
encountering outgoing Hawking modes. The stretched horizon is shown
as the right dashed line, and the global horizon as the left dashed
line. Before the infalling diary scrambles on the stretched horizon,
the outgoing mode is unentangled with it. Only after scrambling will
an infalling observer notice entanglement with the diary. Proper time
along the stretched horizon provides a distinguished set of clocks
which demarcate this interval.}

\end{figure}

There is a finite time delay during which information scrambles on
the stretched horizon after the infalling diary is absorbed. An early
infalling observer (see figure \ref{fig:Different-infalling-observers})
sees no substantial difference from the Unruh or Hartle-Hawking vacua
in this time interval as they cross the global horizon. However an
infalling observer crossing the outgoing mode after this time interval
sees a mode that has had time to spread a distance at least of order
$M$ from the stretched horizon. This mode is now entangled with the
diary.%
\footnote{Note that entanglement depends on the time-slicing in this context,
and is a quantity not accessible to local probes. Nevertheless for
a given time-slicing the state undergoes unitary time evolution and
all external observers agree on the state.%
}

\begin{figure}

\includegraphics[scale=0.6]{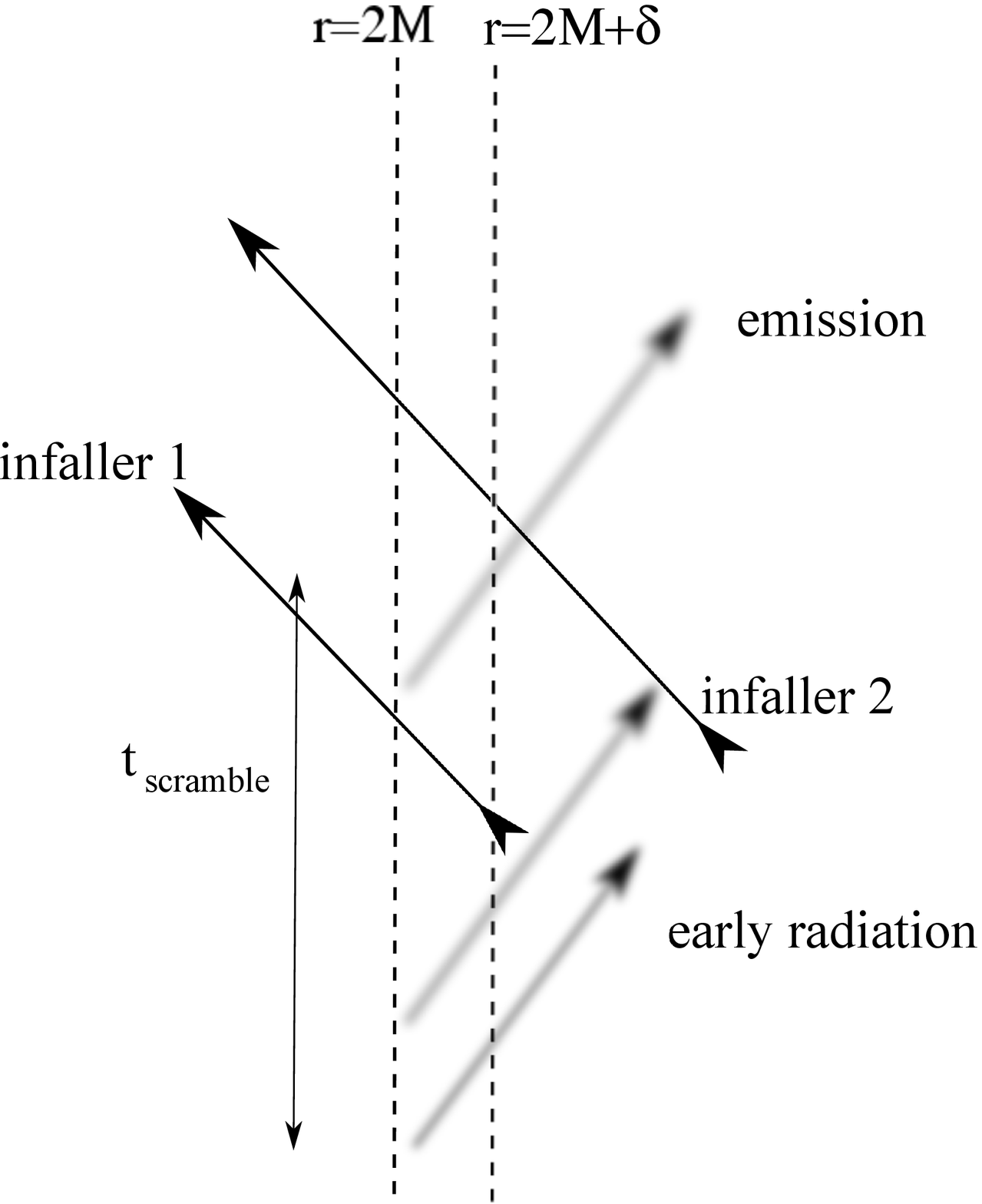}\caption{\label{fig:The-analog-of}The analog of figure \ref{fig:Different-infalling-observers}
with the diary replaced the ordinary Hawking modes. An infaller measuring
a mode outside the stretched horizon (infaller 2) will see it maximally
entangled with the early Hawking radiation. However an earlier infalling
observer (infaller 1) must see vanishing entanglement with the early
Hawking radiation if Postulate 4 holds. }

\end{figure}

Next let us consider the argument of \cite{Almheiri:2012rt} which
essentially replaces the infalling diary by a set of vacuum Hawking
modes. The difference is illustrated in figure \ref{fig:The-analog-of}.
An observer outside the stretched horizon must see entanglement of
the outgoing mode with the early Hawking radiation according to \cite{Hayden:2007cs}.
At the same time, an observer crossing the global horizon sees the
mode entangled with interior modes, not with the early Hawking radiation.
Because the timescale separating infaller 1 and infaller 2 can be
much shorter than the scrambling time, this creates an apparent paradox.
In the work of \cite{Almheiri:2012rt} it is argued that the radiation
on the outside must be some purely outgoing mode at infinity. As noted
above, this can be viewed as a finite excitation of the Boulware vacuum.
The Boulware vacuum has a continuous divergence outside the global
horizon. A freely falling observer will see temperatures of order
the ultraviolet cutoff scale as they cross the stretched horizon.
This then leads to a violation of the postulates of black hole complementarity,
since the physics outside but close to the stretched horizon is no
longer described by a conventional theory.

\begin{figure}

\includegraphics[scale=0.75]{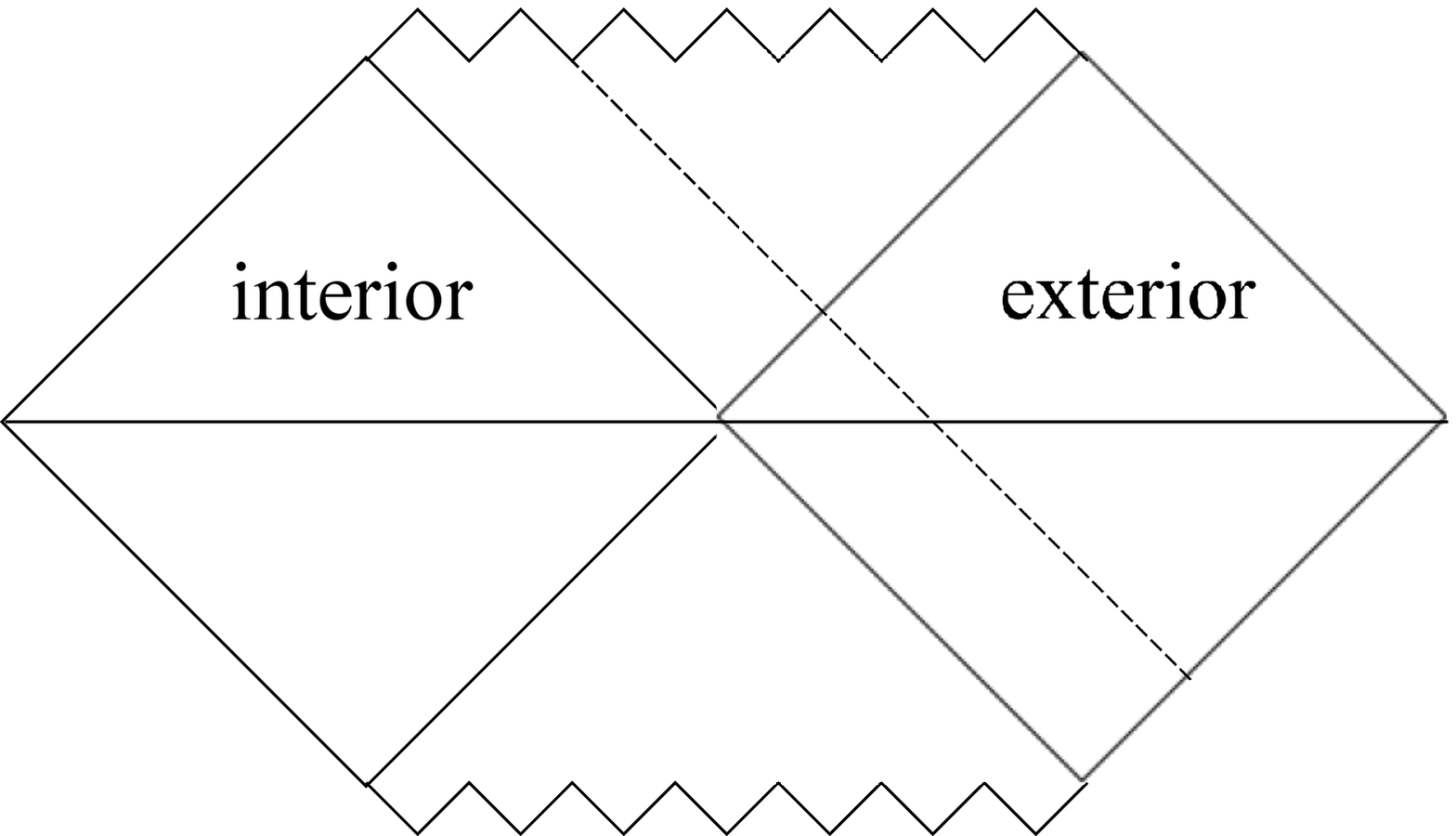}\caption{\label{fig:The-Penrose-diagram}The Penrose diagram for the maximally
extended Schwarzschild black hole. The area to the right of the dashed
line provides a classical model for black hole formation. The Hilbert
space of states may be factored into states on the interior and the
exterior along the time-slice indicated by the horizontal line. Both
sets of modes propagate at later times into the upper quadrant.}

\end{figure}

However it suffices to show the firewall need only ever appear behind
the stretched horizon to exhibit the flaw in the reasoning of \cite{Almheiri:2012rt}.
To do this it is helpful to work with the Hilbert space separated
as shown in figure \ref{fig:The-Penrose-diagram}. The Hartle-Hawking
vacuum involves an entanglement of the modes on each side of the horizon
\cite{Unruh:1976db}. However the left modes never propagate into
the external region on the right. Both sets of modes propagate into
the interior of the black hole, and their entanglement is essential
for the absence of drama for an infalling observer.

In this picture, the exterior modes are a superposition of infalling
and outgoing modes. Consistency with figure \ref{fig:The-analog-of}
then demands that modes representing Hawking particles emitted after
a time of order $ $the Page time $M^{3}$ be maximally entangled
with the earlier radiation. From the exterior viewpoint, there is
no contradiction with unitarity and locality. The problem arises when
one considers an infalling observer. Following the argument of \cite{Almheiri:2012rt}
the exterior mode cannot be simultaneously entangled with the early
Hawking radiation and the interior mode. 

We can model a state where the exterior modes have no entanglement
with the interior modes by simply placing the left interior modes
in their vacuum state. The argument is cleanest if the exterior modes
are placed in a thermal density matrix, rather than a pure state.
Unlike the Boulware vacuum, this does not change the expectation value
of the stress energy tensor in the exterior region %
\footnote{One may also place the exterior modes in a pure state where the expectation
value of the stress energy tensor remains very close to that of the
Hartle-Hawking vacuum. Note such a state involves an infinite number
of infalling and outgoing exterior modes, so the state may be picked
so fluctuations in such a local observable are very small.%
}. It does however produce an infinite firewall on the global horizon.
However this is a crucial difference, because now we have a counterexample
where the firewall only need appear behind the stretched horizon,
and the expectation value of the stress energy tensor need not depart
by a substantial amount from that obtained in the Hartle-Hawking vacuum,
even if the stretched horizon is Planck scale. Therefore we can conclude
that outside a Planck distance from the global horizon there is no
sign that the postulates of black hole complementarity break down.
At or inside the global horizon, all bets are off for a conventional
description of the quantum theory, as emphasized in \cite{Lowe:1995ac,Lowe:1995pu,Lowe:1999pk}.

It is also interesting to estimate whether the outgoing Hawking radiation
leads to an observable deviation in a local quantity outside the stretched
horizon, such as the expectation value of the stress energy tensor.
Along the path of the early infalling observer in figure \ref{fig:Different-infalling-observers}
the result will match that of \cite{Candelas:1980zt}, who found,
for example, a $1/M^{4}$ contribution to the trace of the stress
energy tensor near the horizon due to Hawking radiation in the Hartle-Hawking
vacuum. The later infalling observer will see the same phenomena,
with small differences due to the interference with the earlier outgoing
Hawking radiation. Since the outgoing radiation at this point is maximally
entangled, any fluctuations away from the thermal expectation value
for the stress energy tensor are expected to be down by an extra factor
of $1/M$ or more.

\section{Entropy subadditivity bounds}

Finally, we note that one of the arguments for the firewall of \cite{Almheiri:2012rt}
is based on entropy subadditivity bounds \cite{Araki:1970ba,Lieb:1973cp}.
They divide the system into $A$, the early Hawking modes, $B$ a
late outgoing Hawking mode, and $C$ the interior partner mode of
$B$. They claim the entropy subadditivity bound 
\begin{equation}
S_{AB}+S_{BC}\geq S_{B}+S_{ABC}\,,\label{eq:subadd}
\end{equation}
is violated. Let us analyze this bound, first in the stretched horizon
theory of Postulate 2, and then in a model with timeslices that extend
inside the stretched horizon but terminate on the global horizon.

In the effective field theory of Postulate 2, it is incorrect to view
Hawking radiation as the formation of a maximally entangled pair $B$
and $C$ outside the stretched horizon, with the negative energy $C$
mode subsequently absorbed by the stretched horizon. Introducing $C$
degrees of freedom outside the stretched horizon, and insisting that
they are maximally entangled with the outgoing $B$ modes, indeed
leads to violation of entropy subadditivity as we will see momentarily.
It amounts to cloning of quantum information, and by assumption the
effective field theory of Postulate 2 is a local quantum field theory
where such cloning cannot occur. Rather in the effective field theory
of Postulate 2 only describes the $A$ and the $B$ modes. In this
theory the stretched horizon is a boundary of spacetime. It is a hot
surface from which the Hawking radiation is emitted. At late times
the state of the stretched horizon of the remaining black hole is
maximally entangled with $A$, the early Hawking modes. When a late
Hawking mode $B$ is emitted, the size of the stretched horizon Hilbert
space gets reduced accordingly, and both $B$ and the new stretched
horizon state are separately maximally entangled with $A$. There
is, however, no entanglement between $B$ and the new stretched horizon
state. This is entirely in line with what happens at late times for
burning lump of coal that starts out in a pure state. Entropy subadditivity
reduces to the statement \cite{Araki:1970ba}
\begin{equation}
|S_{A}-S_{B}|\leq S_{AB}\leq S_{A}+S_{B}\label{eq:entropytwo}
\end{equation}
which is close to saturated at late times $S_{AB}=S_{A}-S_{B}$.

We now turn our attention to a description where timeslices extend
inside the stretched horizon and $C$ modes are included. If we take
that description to be a conventional local quantum field theory then
we will run into problems with entropy subadditivity as pointed out
in \cite{Almheiri:2012rt} and these problems can indeed be avoided
by introducing a firewall for infalling observers. It is important
to note, however, that in this case we are no longer considering the
effective field theory of Postulate 2 but have made the further assumption
that local effective field theory can be extended to the region inside
the stretched horizon. 

For the sake of argument, let us instead consider a model where $C$
modes are included and the physics inside the black hole region is
described by some quantum dynamics on the global horizon (rather than
on the stretched horizon as in Postulate 2). A $BC$ pair appears
in a pure state due to a quantum fluctuation and we assume that the
$C$ mode then scrambles with the state on the global horizon in a
time of order $M\log M$. There are two limits where the entropy subadditivity
bound can be easily analyzed, before scrambling has had a chance to
occur, and after the scrambling time. Prior to scrambling, $BC$ remains
in a pure state independent of $A$, so $S_{BC}=0$, and $S_{ABC}=S_{A}+S_{BC}=S_{A}$.
Substituting into \eqref{eq:subadd} yields 
\[
S_{AB}\geq S_{B}+S_{A}\,.
\]
At first sight this seems similar to the analysis of \cite{Almheiri:2012rt}.
However before $C$ is scrambled, we expect the entropy of the outgoing
radiation to increase
\begin{equation}
S_{AB}>S_{A}\,,\label{eq:entropyinc}
\end{equation}
rather than decrease as stated in \cite{Almheiri:2012rt} because
one simply has one additional thermal Hawking particle. We conclude
$S_{AB}=S_{A}+S_{B}$  because $A$ and $B$ are independent prior
to scrambling. 

It is helpful to break the interaction of $C$ with the black hole,
described by some Hilbert subspace $D$, into two steps. First $C$
interacts with the black hole, reducing the number of degrees of freedom
there. This process requires working in some infinite dimensional
Fock space, as is usual in second quantized field theory. However
immediately after this interaction, there will be a dimension $dim(C)$
subspace of the global horizon Hilbert space that is entangled both
with $B$ and with $A$.

After scrambling things become simpler. Scrambling mixes $C$ with
all the other  horizon degrees of freedom. $B$ will become maximally
entangled with $A$, so $S_{AB}=S_{A}-S_{B}$ and the entropy of the
external radiation decreases, $S_{AB}<S_{A}$. After scrambling it
is no longer true that $S_{BD}=0$. Rather if the dimension of $A$
is much larger than $BD$ we expect $B$ and $D$ to become independent,
with $S_{BD}\approx S_{B}+S_{D}=2S_{B}$. Likewise the $BD$ system
will be close to maximally entangled with $A$ after scrambling, so
$S_{ABD}\approx S_{A}-S_{BD}$. Substituting into \eqref{eq:subadd}
yields

\[
S_{A}-S_{B}+S_{BD}\geq S_{B}+S_{A}-S_{BD}\Longleftrightarrow S_{BD}\geq S_{B}
\]
which is satisfied.

It should be noted that the reduced density matrix
\[
\rho_{AB}=\mathrm{Tr_{CD}}\,\rho_{ABCD}
\]
is independent of unitary transformations that act within the $C\times D$
subspace. Therefore the only way to accomplish such a change of entanglement
described above is via a non-unitary transformation. However if we
follow the picture described in the previous section, such a non-unitary
horizon theory is only needed on the global horizon rather than the
stretched horizon, and so remains consistent with the postulates of
black hole complementarity. 

We find no violation of entropy subadditivity implied by the postulates
of black hole complementarity. The correct description of the stretched
horizon theory does not allow for a description of interior $C$ modes
that is independent of the outgoing $B$ modes. By assuming that $BC$
pairs form outside the stretched horizon, \cite{Almheiri:2012rt}
unnecessarily clone information and this leads to the claim that simultaneously
$S_{BC}=0$ (as in our global horizon model prior to scrambling) and
$S_{AB}<S_{A}$ (as in our model only after scrambling).

\section{Discussion}

The main point of this paper is that a firewall for infalling observers
is not an unavoidable consequence of Postulates 1 and 2 as is claimed
by \cite{Almheiri:2012rt}. We do not claim that such a firewall is
impossible. In fact, we have considered several examples where firewalls
do occur. In each case, including that of \cite{Almheiri:2012rt},
the firewall follows from making assumptions about physics in the
region inside the stretched horizon that do not follow from Postulate
2. 

It is interesting to further consider the history of the infalling
observers in figure \ref{fig:Different-infalling-observers}. As a
model for the dynamics inside the horizon, let us imagine we use the
simple non-unitary model described in the previous section. The early
infalling observer, smoothly passes through the stretched horizon
according to Postulate 4. However once an interval of order the scrambling
time passes, the entanglement between the $B$ and the $C$ modes
will change, and this observer will no longer experience a vacuum
state. By this time they will have passed a distance at least of order
$M$ inside the stretched horizon. This, however, coincides with the
location of the curvature singularity. This picture, where information
cloning was prevented by a firewall located near the classical curvature
singularity was advocated in \cite{Lowe:2006xm}, and supported by
AdS/CFT computations in \cite{Lowe:2009mq}. These works focus on
resolving cross-horizon complementarity issues, rather than the outside-the-horizon
issues that are the main focus of the present work. Scrambling therefore
allows the early infalling information to be safely annihilated before
the later infalling observer, who has access to the information from
the exterior Hawking radiation, is able to enter the horizon and potentially
see a contradiction.
\begin{acknowledgments}
We thank J. Polchinski and the authors \cite{Avery:2012tf} for helpful
comments. The research of K.L. and D.A.L. is supported in part by
DOE grant DE-FG02-91ER40688-Task A and an FQXi grant. The research
of L.T. is supported in part by grants from the Icelandic Research
Fund and the University of Iceland Research Fund.
\end{acknowledgments}
\bibliographystyle{apsrev}
\bibliography{firewall}

\end{document}